\documentstyle[preprint,pra,aps]{revtex}
\begin{document}
\title{\bf Relativistic effects in Sr, Dy, YbII and YbIII and search for
variation of the fine structure constant}
\author{V. A. Dzuba, V. V. Flambaum, and M. V. Marchenko}
\address{School of Physics, University of New South Wales,
Sydney 2052,Australia}
\date{\today}
\maketitle

\tightenlines

\begin{abstract}

A possibility for fundamental constants to vary in time is suggested
by theories unifying gravity with other interactions. In this article we
examine proposals to use optical transitions of Sr, Dy, YbII and YbIII for
the search of the time variation of the fine structure constant $\alpha$.
Frequencies of atomic transitions are calculated using relativistic
Hartree-Fock method and configuration interaction technique.
The effect of variation of $\alpha$ on the frequencies is studied by
varying $\alpha$ in computer codes. Accuracy
of measurements needed to improve current best limit on the time
variation of $\alpha$ is discussed.

\end{abstract}
\vspace{1cm}
\pacs{PACS: 06.20.Jr, 31.30.Jv 31.15.Ar}
%*************************************************************************

\section{Introduction}

Theories unifying gravity with other interactions suggest that fundamental
constants could vary in space-time (see, e.g. \cite{vary}). Recent evidence
of variation of the fine structure constant $\alpha$ in quasar absorption
spectra \cite{Webb} elevated interest to the search of variation of $\alpha$
in laboratory experiments. Comparing frequencies of different atomic
transitions over long period of time is a good way to do such search due to
extremely high accuracy of measurements achieved for certain types of
transitions. The best limit on local present-time variation of the fine
structure constant published so far was obtained by comparing Hg$^+$
microwave atomic clock vs hydrogen maser \cite{Prestage}.
Recently this limit was further improved by more than an
order of magnitude in comparing cesium and rubidium atomic clocks
\cite{Pereira}. There are also many proposals for the search of variation
of $\alpha$ in atomic optical transitions, some of which were analyzed
in our previous works (see \cite{optical} and references therein).
In the present paper we analyze three new proposals involving
strontium/calcium, dual beam \cite{Bergeson}, dysprosium atom
\cite{optical,Budker} and ytterbium positive ions $Yb^+$ \cite{Lea}
and  $Yb^{2+}$ \cite{Torgerson}.
We perform relativistic many-body calculations to link variation of
$\alpha$ with the variation of the frequencies of atomic transitions.
Then we use this connection to find out what accuracy
of measurements is needed to improve current best limit on time variation
of the fine structure constant.

In the proposal suggested by S. Bergeson strontium-calcium dual beam is
to be used to compare the frequencies of the $^{1}S_0 - ^{3}P_1$ clock
transitions in these atoms over a long period of time.
Ca and Sr  have similar electron structure. However,
due to higher nuclear charge, relativistic effects are larger for strontium.
If $\alpha$ is changing, corresponding change in frequency of the clock
transition for Sr would go considerably faster than for Ca. Precise
measurements might be able to indicate this or, at least, put strong
constrain on possible variation of $\alpha$. Calculations of the
relativistic effects for Ca were done in our previous work \cite{optical}.
In present paper we do similar calculations for Sr.

Experiments with ytterbium positive ion have advantages of greater
relativistic effects due to larger nuclear charge and the convenience
of working with two different transitions of the same element.
There are two transitions in Yb$^+$ involving metastable states for
which comparison of frequencies is considered. One is quadrupole transition
$4f^{14}6s \  ^2S_{1/2} - 4f^{14}5d  \ ^2D_{5/2}$ and another is octupole
transition $4f^{14}6s  \ ^2S_{1/2} - 4f^{13}6s^{2} \ ^2F_{7/2}$.
The quadrupole transition is basically a $s - d$ transition while
the octupole one is a $f - s$ transition.
According to simple analytical formula presented in Ref. \cite{optical}
relativistic energy shifts for $s$ electrons, and electrons
with high total momentum $j$ (like $d$ and $f$ electrons) are large
but have opposite sign.
This means that we should expect that two metastable states of
Yb$^+$ move in opposite directions if $\alpha$ is changing. This
brings extra enhancement to the sensitivity of the measurements for
Yb$^+$ to the variation of $\alpha$. Our accurate calculations presented
below support these considerations.

The proposal for dysprosium is quite different from what was considered
so far. Instead of comparing two very stable atomic clock frequencies
the authors of this proposal \cite{optical} suggest to measure very
small frequency of the transition between two
almost degenerate states of opposite parity in dysprosium. The states are
$4f^{10}5d6s  \ ^3[10]_{10} \ \mbox{E}=19797.96\mbox{cm}^{-1}$  and
$4f^95d^26s \  ^9K_{10} \ \mbox{E}=19797.96\mbox{cm}^{-1}$.
These states were used before for the search of parity
non-conservation in Dy \cite{BudkerPNC}. Small energy splitting
and different electron structure of these two states lead to very
strong enhancement of the sensitivity of the frequency of transition
between the states to variation of $\alpha$. The enhancement (about
eight orders of magnitude) seems to be strong enough to overcome
the disadvantage of dealing with states which are not very narrow.

In the present paper we calculate the values of relativistic energy
shifts for Sr, Yb$^+$ and Dy and discuss what accuracy of measurements
is needed to improve current best constrain on local time variation
of the fine structure constant.

\section{Calculations of energies}

We use relativistic Hartree-Fock (RHF) and configuration interaction (CI)
methods to do the calculations.

RHF Hamiltonian is used to generate a set of single-electron orbitals.
We use a form of singe-electron wave function which explicitly depends
on the fine structure constant $\alpha$:
\begin{equation}
	\psi(r)_{njlm}=\frac{1}{r}\left(\begin {array}{c}
	f_{n}(r)\Omega(\mathbf{n})_{\mathit{jlm}}  \\
	i\alpha g_{n}(r)  \tilde{ \Omega}(\mathbf{n})_{\mathit{jlm}}
	\end{array} \right)
\end{equation}

Then the RHF equation for $\psi(r)_n$ has the following form
(in atomic units)
\begin{equation}
	\begin {array}{c} f^{'}_n(r)+\frac{\kappa_{n}}{r}f_n(r)-
	[2+\alpha^{2}(\epsilon_{n}-\hat{V}_{HF})]g_n(r)=0,  \\
	g^{'}_n(r)-\frac{\kappa_{n}}{r}f_n(r)+(\epsilon_{n}-
	\hat{V}_{HF})f_n(r)=0, \end{array} \label{Dirac}
\end{equation}
here $\kappa=(-1)^{l+j+1/2}(j+1/2)$, $n$ is the principal quantum number and $\hat{V}_{HF}$ is the
Hartree-Fock potential.
The value of relativistic effects is studied by varying the value
of $\alpha$ in (\ref{Dirac}). In particular, non-relativistic limit
corresponds to $\alpha = 0$.

In the CI calculations we use approach similar to what was first developed
in Ref. \cite{CI+MBPT}. Electrons occupying open shells are considered as
valence electrons and all other are core electrons. Correlations between
valence electrons are treated within the CI method while correlations
between valence and core electrons are included by means of the
many-body perturbation theory.
The effective CI Hamiltonian for valence electrons is presented by
\begin{equation}
	\hat H^{CI}= \sum^{N}_{i=1} \hat h_{i} + \sum_{i<j} \frac{e^2}{r_{ij}}.
\end{equation}
Here $N$ is the number of valence electrons and $\hat h_i$ is an effective
singe-electron Hamiltonian
\begin{equation}
	\hat h_{i}= c \bbox{\alpha p} +(\beta -1)mc^2 - \frac{Ze^2}{r_{i}}
	+ \hat{V}_{\rm core} +\hat{\Sigma}_{1}
\label{hi}
\end{equation}
Here $V_{\rm core}$ is the Hartree-Fock potential created by core electrons.
It differs from $V_{HF}$ in Eq. (\ref{Dirac}) by contribution of valence
electrons. $\hat \Sigma_1$ is so called ``correlation potential'' operator.
It describes correlations between a particular valence electron and
core electrons (see Ref. \cite{CI+MBPT} for details). Note that
in contrast with Ref. \cite{CI+MBPT} we don't include in present work
the $\hat \Sigma_2$ operator, which is a two-electron operator describing different type of correlations between valence and core electrons.
Terms with $\hat \Sigma_2$ can be considered as screening of Coulomb
interaction between valence electrons by core electrons.
These terms are less important than those with $\hat \Sigma_1$ but much
more time consuming in calculations.
We either neglect them or simulate their effect by
introducing screening factors.

We are now going to discuss the specifics of the calculations for each
atom/ion. Apart from the states of interest we also calculate energies of
the other states of the same configurations to ensure that the accuracy
is systematically good. We also calculate magnetic $g$-factors to ensure
correct identification of states. This is particularly important for dysprosium.

\subsection{Strontium}

Strontium in its ground state is a closed-shell atom. It has two
$5s$-electrons on its outermost shell and we need to consider energy
intervals between $^1S_0$ ground state and states of the $5s5p$ configuration
where the $^3P_1$ metastable state is of most interest.
The RHF calculations for Sr were done in $V^N$ approximation, for a
closed-shell atom in its ground state. For the CI calculations we considered
Sr as an atom with two valence electrons and followed the similar
calculations for Ba \cite{DzubaBa}. Basis states for the CI+MBPT method were
calculated using the B-spline technique \cite{Johnson86} with 40 B-splines
in a cavity of radius $R=40 a_B$. The same basis functions were used
to calculate $\hat \Sigma_1$ and for the CI calculations. Thirteen lowest
states above core in each of the $s_{1/2}$, $p_{1/2}$, $p_{3/2}$, $d_{3/2}$
and $d_{5/2}$ waves were used to construct two-electron wave function
for both $5s^2$ and $5s5p$ configurations. Large number of basis functions
is needed mostly for adequate description of the $5s5p$ configuration.
This is because the $V^N$ approximation doesn't provide us with a good $5p$
single-electron state.
Also, the $5s$ single-electron state in the $5s5p$ configuration is
different from the $5s$ state in the $5s^2$ configuration for which
Hartree-Fock calculations  were done. However, with thirteen states in
each wave the saturation of the basis was clearly achieved and adding
more states to the basis didn't change the energy. Two-electron basis
states for the CI calculations were obtained by distributing valence
electrons over 65 basis states ($13 \times 5$) in all possible ways with
a restriction of fixed parity and total momentum.

The results are presented in
Table \ref{energy}. As one can see the accuracy for the state of the
 interest $^3P_1$ is better than 1\% while accuracy for other
states is also good.

\subsection{Ytterbium}

The ground state of ytterbium positive ion is $4f^{14}6s \ ^2S_{1/2}$
and we need to consider transitions into the $4f^{14}5d \ ^2D_{5/2}$ and
$4f^{13}6s^2 \ ^2F_{7/2}$ states. Therefore it is convenient to do the
RHF calculations in the $V^{N-1}$ approximation, for the Yb$^{2+}$ ion
with the $4f^{14}$ closed-shell configuration. The $6s$, $5d$ and other
basis states for the CI method are calculated then in the field of
frozen closed-shell core of Yb$^{2+}$. Then, in the CI calculations, we
need to consider all $4f$ electrons as valence ones since one of the
transitions of the interest involves excitation from the $4f$ subshell.
So, the total number of valence electrons in present CI calculations is
fifteen.
This is very different from our previous calculations for Yb$^+$
\cite{optical} in which the $4f^{13}6s^2 \ ^2F_{7/2}$ state was not
considered and we were able to treat ytterbium ion as a system with
one external electron above closed shells.

Our final set of single-electron states for the CI calculations consisted
of $4f_{5/2}$, $4f_{7/2}$, $6s_{1/2}$, $5d_{3/2}$, $5d_{5/2}$ and few more
$s$ and $f$ states above $4f$ and $6s$. Note that in contrast with Sr
we don't need many basis functions here because all our single-electron
wave functions correspond to the Yb$^+$. This makes initial approximation
to be very good and leads to fast convergence of the CI calculations with
respect to the basis set used.

We also don't include $\hat \Sigma_1$ in calculations for Yb$^+$.
In a case of many valence electrons (fifteen for Yb$^+$) correlations
are dominated by correlations between them which are taken into account
accurately via the CI technique. Correlations between valence electrons
and core electrons mostly manifest themself via screening of the Coulomb
interaction between valence electrons. We take this effect into account
semiempirically, by introducing screening factors $f_k$. Namely, we
multiply every Coulomb integral of the multipolarity $k$ by a numerical
factor $f_k$ which is chosen to fit the energies. It turns out that good
fit for Yb$^+$ is achieved with $f_2 =0.8$ and $f_k =1$ for all other $k$.

Many-electron basis states for the CI calculations were obtained by
allowing all possible single and double excitations from the base
configuration with the restriction of fixed parity and total momentum.

Results for energies of Yb$^+$ are presented in Table \ref{energy}.
The theoretical accuracy for energies as compared to the experiment
is 2- 3\% for the states of  interest and is not worse than 5\%
for other states.

\subsection{Dysprosium}

Dysprosium atom is the most difficult for calculations because of its
complicated electron structure. Ground state configuration of Dy is
$4f^{10}6s^2$ which means that there is no realistic RHF approximation
which corresponds to a closed-shell system. We do the RHF calculations
for Dy in the $V^N$ approximation with an open-shell version of
the RHF method.
Contribution of the $4f$ electrons into the RHF potential is calculated as
for a closed shell and then multiplied by a numerical factor to take
into account its fractional occupancy. This factor is 10/14 when interaction
of the $4f$ electrons with other core electrons is considered and 9/13
when interaction of a $4f$ electron with other $4f$ electrons is considered.
When convergence is achieved we have the $4f$ and $6s$ basis states for the
CI calculations. To calculate other states of valence electrons we remove
one $6s$ electron, freeze all RHF orbitals, including $4f$ and $6s$ and
calculate the $6p_{1/2}$, $6p_{3/2}$, $5d_{3/2}$, $5d_{5/2}$
and few more $d$-states above $5d$ in the field
of frozen RHF core.

In the CI calculations states below $4f$ are considered as core states
and all other as valence states. Total number of valence electrons is
therefore twelve. As for the case of Yb$^+$ we neglect $\hat \Sigma_1$
and use screening factors as fitting parameters to improve agreement
with experiment. It turns out that best fit for the $4f^{10}6s6p$
configuration is achieved with $f_1=0.7$ and $f_k=1$ for all other $k$.
No fitting was used for other configurations.

To calculate states of the $4f^{10}6s^2$, $4f^{10}6s6p$ and $4f^{10}6s5d$
configurations we use the $4f_{5/2}$, $4f_{7/2}$, $6s_{1/2}$, $6p_{1/2}$,
$6p_{3/2}$, $5d_{3/2}$ and $5d_{5/2}$ single-electron basis
functions and all possible configurations which can be obtained from these
basis functions by exciting of one or two electrons from the base
configuration. Same approach doesn't work for the $4f^9 5d^2 6s$
configuration because of huge number of many-electron basis states
generated this way and as a consequence, the CI matrix is of so large size
that it could not be handled by our computers. On the other hand test
 calculations
with pairs of configurations showed that mixing of our state of interest
with other configurations is small and can be neglected. We do need
however to include mixing with the $4f^9 5d6d 6s$, $4f^9 5d7d 6s$ and
$4f^9 6d^2 6s$ configurations.
This is because our basis $5d$ state corresponds rather
to the $4f^{10} 5d 6s$ configuration and extra $d$-states are needed
to correct it.

The result are presented in Table \ref{energy}. Note that they are
considerably better than in our previous calculations \cite{DzubaDy}.
This is because of better basis and more complete CI treatment.

\section{Frequency shift: results and discussions}

In the vicinity of the physical value of the fine structure constant
($\alpha=\alpha_{0}$) frequency ($\omega$) of an atomic transition can
be presented in a form
\begin{equation}
	\omega=\omega_{0}+qx ,
\label{omega}
\end{equation}
where $ x = (\alpha^{2}/\alpha^{2}_{0})-1$,
$\omega_{0}$ is the experimental value of the frequency and $q$ is
a coefficient which determines the frequency dependence on the
variation of $\alpha$. To find the values of $q$ for different atomic
transitions we repeat all calculations for
$\alpha = \sqrt{\frac{9}{8}}\alpha_0$, $\alpha = \alpha_0$ and
$\alpha = \sqrt{\frac{7}{8}}\alpha_0$. Then
\begin{equation}
	q=4(\omega_{+}-\omega_{-}),
\end{equation}
Where $\omega_{+}$ is the value of $\omega$ for
$\alpha = \sqrt{\frac{9}{8}}\alpha_0$, and $\omega_{-}$
is the value of $\omega$ for$\alpha = \sqrt{\frac{7}{8}}\alpha_0$.
Calculations for  $\alpha=\alpha_0$ are done to compare the theory
with experiment and to check whether frequencies are linear functions
of $\alpha^2$. The results for coefficients $q$ are presented in Table
\ref{q-s}. Note that we have included in the Table the results of our old
calculations for Yb$^+$. These calculations were done in a very different
way, assuming that Yb$^+$ is an atom with one external electron above closed
shells.
Comparison of the results obtained by different methods
gives estimate of the accuracy of calculations.

Search for the time variation of the fine structure constant can
be conducted by comparing two   frequencies of atomic transitions
over long period of time. The measured value can be presented as
\cite{optical,Prestage}
\begin{equation}
	\Delta(t)=\left( \frac{\dot \omega_1}{\omega_1} -
	\frac{\dot \omega_2}{\omega_2} \right).
\label{rate}
\end{equation}
Using Eq. (\ref{omega}) one can reduce Eq. (\ref{rate}) to
\begin{equation}
	\Delta(t)=\left( \frac{2q_1}{\omega_1} -
	\frac{2q_2}{\omega_2} \right)(\frac{\dot \alpha}{\alpha_0}).
\label{rateq}
\end{equation}
Current best laboratory limit on the time variation of $\alpha$ is
$\dot \alpha/\alpha < 10^{-15}\mbox{yr}^{-1}$ \cite{Pereira}.

In the first experiment considered in this paper a dual calcium-strontium
beam is to be used to compare the frequencies of the $^1S_0 - ^3P_1$
transitions in both atoms. Substituting $\omega_1=15210 \mbox{cm}^{-1}$,
 $q_1=230 \mbox{cm}^{-1}$ for Ca \cite{optical},
$\omega_2=14504 \mbox{cm}^{-1}$,
 $q_2=667 \mbox{cm}^{-1}$ for Sr (Tables \ref{energy},\ref{q-s}) and
$\dot \alpha/\alpha = 10^{-15}\mbox{yr}^{-1}$ we get
\begin{equation}
	\Delta(t)(\mbox{Sr-Ca})=  6.2 \times 10^{-17} \mbox{yr}^{-1}.
\label{Sr-Ca}
\end{equation}
Note that the width of $ ^3P_1$ state in Sr may be a problem in this case.

In the case of Yb$^+$ frequencies of the $^2S_{1/2} - ^2D_{5/2}$ and
$^2S_{1/2} - ^2F_{7/2}$ are to be compared. Substituting the numbers we
get
\begin{equation}
	\Delta(t)(\mbox{Yb}^+)=  6.1 \times 10^{-15} \mbox{yr}^{-1}.
\label{Yb+}
\end{equation}
Note two orders of magnitude improvement in the magnitude in comparison with
the Sr-Ca dual beam experiment.

  We have also calculated $q$-coefficient for
    $4f^{14}$ $^1S_{0}$ -  $4f^{13}5d$  $^3P_{0}$
 ($\omega=45276$ cm$^{-1}$) transition from  YbIII ground state.
This was motivated by the proposed measurements \cite{Torgerson}
 of $\alpha$-variation
using comparison of  $^1S_{0}$ - $^3P_{0}$ transition frequencies
in In$^+$,  Tl$^+$  and odd isotope of Yb$^{++}$.
 The different signs and magnitudes
of relativistic corrections in In$^+$ ($q$= 4414 cm$^{-1}$), Tl$^+$
($q$=19745  cm$^{-1}$) and  Yb$^{++}$   ($q$= -27800 cm$^{-1}$)
provide an excellent control of systematic errors
since systematic errors are not correlated with signs and magnitudes of
the frequency shifts $q x$, where $ x = (\alpha^{2}/\alpha^{2}_{0})-1$.
The same idea (combination of anchors, positive shifters and negative
 shifters) has been used to control systematic errors in Ref. \cite{Webb}.

In our view, a very interesting possibility is that  for dysprosium.  Instead
of comparing frequencies of different transitions one should measure
the energy difference between two very close states of opposite
parity. The corresponding $q$-coefficient is
$q = 6008 + 23708= 29716\mbox{cm}^{-1}$ (see Table \ref{q-s}).
The frequency of this transition ranges from few MHz to few GHz depending
on isotopes and hfs components used. If we take, e.g.
 $\omega=3.1 \mbox{MHz}$ \cite{BudkerDy} we get
\begin{equation}
	\Delta(t)(\mbox{Dy})=  5.7 \times 10^8 (\frac{\dot \alpha}{\alpha_0}).
\label{Dy1}
\end{equation}
This is an eight orders of magnitude enhancement in the relative value
of the effect compared
to atomic clock transitions!
Substituting
$\dot \alpha/\alpha = 10^{-15}\mbox{yr}^{-1}$ we get
\begin{equation}
	\Delta(t)(\mbox{Dy})=  5.7 \times 10^{-7} \mbox{yr}^{-1}.
\label{Dy2}
\end{equation}
This means that to improve current best limit on local time variation of
$\alpha$ the frequency of this transition in Dy should be measured
to the accuracy of about $10^{-7}$ over about a year time interval.
This seems to be feasible \cite{Budker}.

\section{ Acknowledgments}

We are grateful S. Bergeson, P. Blythe, D. Budker,S. Lea, F. Pereira
and J.R. Torgerson for useful discussions.
Part of the work was done on the computers of the
Australian Center for Advanced Computing and Communications.
This work is supported by the Australian Research Council.
%####################################################################

%####################################################################
\begin{table}
\caption{Energies and g-factors}
\label{energy}
\begin{tabular}{ccllrrrllrlcc}
 \multicolumn{2}{c}{Atom/ion}& \multicolumn{2}{c}{State} &   \multicolumn{2}{c}{Energy, 1/cm }&\multicolumn{2}{c}{g }\\&&&&theory & experiment\tablenotemark[1]
  & theory   &experiment\tablenotemark[1]  &non-relativistic&
  \\
\hline
&  $\rm{Sr}$      & 5$s^2$             & $^1S_0$         & 0        &  0           &               &       &     \\
&                 & 5$s5p$             & $^3P_0$         & 14171    &  14318       &               &       &     \\
&                 &5$s5p$              & $^3P_1$         & 14384    &   14504      &               &       &     \\
&                 &5$s5p$              & $^3P_2$         & 14832    &  14899       &               &       &     \\
&                 &5$s5p$              & $^1P_1$         &  22829   &  21698       &               &       &     \\
&  $\rm{Dy}$      &4$f^{10}$6$s^2$     &$^5I_8$          & 0        &   0          &   1.243       &1.242  &1.25     \\
&                 &4$f^{10}$6$s^{2}$   &$^5I_7$          &4123      &  4134        &  1.175        & 1.173 &1.179     \\
&                 &4$f^{10}$6$s^{2}$   &$^5I_6$          &7147      &  7051        &   1.072       & 1.072 &1.071     \\
&                 &4$f^{10}$6$s^{2}$   &$^5I_5$          &9428      &  9212        &   0.907       & 0.911 &0.9     \\
&                 &4$f^{10}$6$s^{2}$   &$^5I_4$          &11199     &  10925       &   0.614       & 0.618 &0.6     \\
&                 &4$f^{10}$5$d $6$s$  &$^3[8]_9$        &18605     &  17515       &   1.319       & 1.316 &     \\
&                 &4$f^{10}$5$d$6$s$   &$^3[9]_{10}$     & 18615    &   18463      &   1.291       &1.282  &     \\
&                 &4$f^{10}$5$d$6$s$   & $^3[10]_{11}$   &  19811   &  19349       &  1.268        &1.27   &     \\
&                 &4$f^{10}$5$d$6$s$   & $^3[10]_{10}$   & 20133    &   19798      &   1.208       &1.21   &     \\
&                 & 4$f^{9}$5$d^2$6$s$ & $^9K_{12}$      & 23345    & 22541        &  1.327        &1.333  &1.333     \\
&                 & 4$f^{9}$5$d^2$6$s$ & $^9K_{11}$      &20513     &  20448       &   1.352       &1.354  &1.303     \\
&                 & 4$f^{9}$5$d^2$6$s$ & $^9K_{10}$      & 19623    &  19798       &   1.372       &1.367  &1.264     \\
&                 & 4$f^{9}$5$d^2$6$s$ & $^9I_9$         & 19434    & 19558        &   1.390       &1.39   &1.377     \\
&                 & 4$f^{9}$5$d^2$6$s$ & $^9G_8$         & 18379    &  18473       &  1.461        &1.46   &1.5     \\
&                 & 4$f^{9}$5$d^2$6$s$ & $^9G_7$         & 18662    & 18529        &   1.492       & 1.467 &1.5     \\
&                 & 4$f^{9}$5$d^2$6$s$ & $^7F_6$         &  19714   &  19304       &   1.527       & 1.54  &1.5     \\
&                 & 4$f^{9}$5$d^2$6$s$ & $^7G_5$         &  21697   & 20892        &    1.510      &1.32   &1.37     \\
&                 & 4$f^{9}$5$d^2$6$s$ & $^9G_4$         &  23748   &  22697       &    1.492      & 1.487 &1.5     \\
&  $\rm{Yb}^+$    &4$f^{14}$6$s$       &$^2S_{1/2}$      &   0      &  0           &  2.000        &1.998  &2     \\
&                 &4$f^{14}$5$d$       &$^2D_{3/2}$      &   22888  & 22961        &  0.800        &0.800  &0.8     \\
&                 &4$f^{14}$5$d$       & $^2D_{5/2}$     & 23549    & 24333        &  1.200        &1.202  &1.2     \\
&                 &4$f^{13}$6$s^2$     &$^2F_{5/2}$      &31820     & 31568        &  0.857        &0.862  &0.857     \\
&                 &4$f^{13}$6$s^2$     & $^2F_{7/2}$     &   21819  & 21419        &  1.143        &1.145  &1.143     \\
&                 &4$f^{14}$6$p$       & $^2P_{1/2}$     &26000     & 27062        &  0.667        &0.667  &0.667     \\
&                 &4$f^{14}$6$p$       &$^2P_{3/2}$      &29005     & 30392        &  1.333        &1.333  &1.333
\end{tabular}
\tablenotetext[1]{References \cite{Moore,Dy}}

\end{table}

\begin{table}
\caption{Relativistic energy shift $q$ (cm$^{-1}$).}
\label{q-s}
\begin{tabular}{lllrr}
 \multicolumn{1}{c}{Atom/ion} & \multicolumn{2}{c}{State}  & This work  &
\cite{optical}  \\
\hline
 Sr     & $5s5p$          & $^3P_1$        &    667 &     \\
        & $5s5p$          & $^1P_1$        &   1058 &     \\
 Dy     & $4f^{10}5d6s$   & $^3[10]_{10}$  &   6008 &      \\
        & $4f^{9}5d^26s$  & $^9K_{10}$     & -23708 &       \\
 YbII   & $4f^{14}5d$     & $^2D_{3/2}$    &  10118 & 12582  \\
        & $4f^{14}5d$     & $^2D_{5/2}$    &  10397 & 11438  \\
        & $4f^{13}6s^2$   & $^2F_{7/2}$    & -56737 &       \\
 YbIII  & $4f^{13}5d$     & $^3P_{0}$      & -27800 &        \\
\end{tabular}
\end{table}

\end{document}